\documentclass{elsart}

\begin{document}

\begin{frontmatter}

\title{A new approach to the epsilon expansion of generalized hypergeometric functions}

\author[zgz,Ann]{David Greynat}  and \author[zgz]{Javier Sesma}

\address[zgz]{Departamento de F\'{\i}sica Te\'orica,
         Facultad de Ciencias, Universidad de Zaragoza, 50009 Zaragoza, Spain}

\address[Ann]{LAPTh., Univ. de Savoie, CNRS, B.P. 110, Annecy-le-Vieux F-74941, France}

\begin{abstract}
Assuming that the parameters of a generalized hypergeometric function depend linearly on a small variable $\varepsilon$, the successive derivatives of
the function with respect to that small variable are evaluated at $\varepsilon=0$ to obtain the coefficients of the $\varepsilon$-expansion of the
function. The procedure, which is quite naive, benefits from simple explicit expressions of the derivatives, to any order, of the Pochhammer and reciprocal
Pochhammer symbols with respect to their argument. The algorithm may be used algebraically, irrespective of the values of the parameters. It reproduces the exact results obtained by other authors in cases of especially simple parameters. Implemented numerically, the procedure improves considerably, for higher orders in $\varepsilon$, the numerical expansions given by other methods.
\end{abstract}


\end{frontmatter}

\begin{tabbing}
{\it Corresponding author:} \quad \=  David Greynat \\
{\it address:} \> LAPTh.\+ \\
                  Univ. de Savoie, CNRS,\\
                  B.P.110, Annecy-le-Vieux F-74941, France \- \\
{\it phone:} \> 33 450 091 692\\
{\it fax:}   \> 33 450 098 913\\
{\it e-mail:} \> david.greynat@gmail.com
\end{tabbing}

\section{Introduction}

Special functions of the hypergeometric class have experienced a considerable interest in the last decade due to their connection with Feynman integrals
in quantum field theory. Recurrence relations among hypergeometric functions allow one to reduce the number of Feynman integrals to be computed in a
given process. In reciprocity, calculations with Feynman integrals have revealed new relations between hypergeometric functions \cite{knie}.

Dimensional regularization of Feynman integrals can be conveniently carried out if one knows the $\varepsilon$-expansion of the related generalized
hypergeometric functions, that is, an expansion in powers of a small variable $\varepsilon$ on which the parameters of the function depend linearly.
Several procedures have been suggested to deal with this issue. Algebraic methods applicable in the cases of integer, half-integer or rational
parameters \cite{dav1,dav2,dav3,jege,dav4,kalm,kal1,kal2,kal3,yost} lead to expansions whose coefficients can be written in terms of generalized polylogarithms.
Computer packages implementing algorithms based on the Hopf algebra of nested sums \cite{moc1,wei1} are available. Let us mention, for instance, a C++
program \cite{wei2} in the framework of GiNaC, \textsc{XSummer} \cite{moc2}, which uses the computer algebra system \textsc{Form}, \verb"HypExp" and
\verb"HypExp" 2 \cite{hub1,hub2}, based on \verb"Mathematica", and \verb"HYPERDIRE" \cite{byte}, useful for differential reduction of hypergeometric
functions.
A recently proposed package, \verb"NumExp"  \cite{huan}, with a different strategy, allows one to get the $\varepsilon$-expansion as a Laurent series
whose coefficients are evaluated numerically by using a multi-precision finite-difference method.

Here we are concerned with the $\varepsilon$-expansion, to all orders, of generalized hypergeometric functions of one variable, $\ _p\! F_q $, or of
several variables, such as the Appell  and  the Kamp\'e de F\'eriet functions  \cite{Appell}, with parameters depending linearly on $\varepsilon$. No
restriction is placed on the parameters, which may be complex.
Our procedure stems from the well-known series expansions of those functions and, similarly to the method in Ref. \cite{huan}, consists in obtaining
the coefficients of the Laurent series by multiple derivation with respect to $\varepsilon$ followed by particularization for $\varepsilon=0$. Instead
of using (approximate) numerical derivation, like in Ref. \cite{huan}, we benefit from simple formulae for the derivatives, of any order, of the
Pochhammer and reciprocal Pochhammer symbols, which allow us to obtain explicit algebraic expressions for the successive terms of the
$\varepsilon$-expansion.
The validity of our expressions is subordinate to the convergence of the series expansions of the functions considered. We assume in what follows that
the values of the variables are such that the required convergence is guaranteed. Otherwise, such expressions have a purely formal character. Needless to
say, analytic continuation allows one to cover a wider range of values of the variables.

We present our algorithm in Section 2. For the sake of clarity, we consider the $\varepsilon$-expansion of
\begin{equation}
\ _p\! F_q \left( \!\! \left.\begin{array}{l}\alpha_1,\ldots,\alpha_{p}\\\beta_1,\ldots,\beta_q\end{array}\right|z\right) \label{i1}
\end{equation}
with upper and lower parameters of the form
\begin{equation}
\alpha_i=A_i+a_i\,\varepsilon ,\quad i=1, \dots , p,   \qquad \beta_j=B_j+b_j\,\varepsilon, \quad j=1, \ldots , q,  \label{i2}
\end{equation}
without any restriction on the values of $A_i$, $a_i$, $B_j$, $b_j$. The procedure can be trivially extended to Appell and Kamp\'e de F\'eriet
functions. By way of illustration, we report in Section 3 the results of applying our method to several examples of generalized hypergeometric
functions found in the literature. Section 4 contains some pertinent comments. We have devoted Appendices A and B to providing expressions for the
derivatives of the Pochhammer and reciprocal Pochhammer symbols.

\section{The algorithm}

Our starting point, as stated earlier, is the series expansion
\begin{equation}
\ _p\! F_q \left( \!\! \left.\begin{array}{l}\alpha_1,\dots,\alpha_{p}\\
\beta_1,\ldots,\beta_q\end{array}\right|z\right) = \sum_{m=0}^\infty\,\frac{(\alpha_1)_m \cdots(\alpha_p)_m}{(\beta_1)_m
\cdots(\beta_q)_m}\,\frac{z^m}{m!}\,, \label{ii1}
\end{equation}
where we use the Pochhammer symbols, $(\alpha)_n \doteq \Gamma(\alpha+n) /  \Gamma(\alpha)$.

We say that all lower parameters $\beta_j$ are {\em regular} if all $B_j$ ($j=1, \ldots, q$) are different from zero or a negative integer. In this
case the $\varepsilon$-expansion is a Taylor one, free of negative powers of $\varepsilon$. If one or several lower parameters are {\em singular}, that
is, if some of the $B_j$ are zero or a negative integer, negative powers of $\varepsilon$ may be present and the expansion is a Laurent one. We treat
 the two cases in turn.

\subsection{All lower parameters are regular}

Let us denote by $\mathcal{C}_n(z)$ the coefficients of the $\varepsilon$-expansion
\begin{equation}
\ _p\! F_q \left( \!\! \left.\begin{array}{l}\alpha_1,\ldots,\alpha_{p}\\
\beta_1,\ldots,\beta_q\end{array}\right|z\right) = \sum_{n=0}^\infty\,\mathcal{C}_n(z)\,\varepsilon^n\,. \label{ii2}
\end{equation}
Obviously,
\begin{equation}
\mathcal{C}_n(z) = \frac{1}{n!}\left.\frac{\partial^n}{\partial \varepsilon^n}\ _p\! F_q \left( \!\!
\left.\begin{array}{l}\alpha_1,\ldots,\alpha_{p}\\
\beta_1,\ldots,\beta_q\end{array}\right|z\right)\right|_{\varepsilon=0}.  \label{ii3}
\end{equation}
The hypergeometric function $\ _p\! F_q$ is given, in Eq. (\ref{ii1}), as a series of products of functions of $\varepsilon$ of the form
\[
\prod_{l=1}^{p+q}\,f_{l,m}(\varepsilon)\,\frac{z^m}{m!}\,,  \qquad m=0, 1, 2, \ldots\,,
\]
with
\[
f_{i,m}(\varepsilon)\equiv (\alpha_i)_m\,,\qquad i=1, 2, \ldots, p\,,
\]
\[
f_{j,m}(\varepsilon)\equiv \frac{1}{(\beta_j)_m}\,,\qquad j=p+1, p+2, \ldots, p+q\,.
\]
The well-known Leibniz formula for the $n$th derivative of  a product of two functions,
\[
\frac{d^n}{d\varepsilon^n}\,f(\varepsilon)\,g(\varepsilon) = \sum_{k=0}^n {n \choose k}\, \left(\frac{d^k}{d\varepsilon^k}\,f(\varepsilon)\right)
\left(\frac{d^{n-k}}{d\varepsilon^{n-k}}\,g(\varepsilon)\right)\,,
\]
can be trivially extended to get
\begin{eqnarray}
\frac{d^n}{d\varepsilon^n}\,\prod_{l=1}^{p+q}\,f_{l}(\varepsilon) &=& \sum_{k_1=0}^n {n \choose k_1} \left(\frac{d^{k_1}}{d\varepsilon^{k_1}}\,
f_1(\varepsilon)\right)\, \sum_{k_2=0}^{n_1} {n_1 \choose k_2} \left(\frac{d^{k_2}}{d\varepsilon^{k_2}}\,f_2(\varepsilon)\right)\,\cdots \nonumber  \\
& & \hspace{-70pt}\cdots\,\sum_{k_{p+q-1}=0}^{n_{p+q-2}} {n_{p+q-2} \choose k_{p+q-1}} \left(\frac{d^{k_{p+q-1}}}{d\varepsilon^{k_{p+q-1}}}\,
f_{p+q-1}(\varepsilon)\right)\,\left(\frac{d^{n_{p+q-1}}}{d\varepsilon^{n_{p+q-1}}}\,f_{p+q}(\varepsilon)\right)\,,  \label{ii3bis}
\end{eqnarray}
where we have denoted
\begin{equation}
n_l=n-k_1-k_2-\ldots-k_l, \qquad (l=1, 2, \dots ,p+q-1).   \label{ii6}
\end{equation}

Let us introduce the abbreviations
\begin{equation}
\mathcal{P}_m^{(k)}(\alpha) \equiv \frac{1}{k!}\,\frac{d^k}{d\alpha^k}(\alpha)_m,  \qquad
\mathcal{Q}_{m}^{(k)}(\beta) \equiv \frac{1}{k!}\,\frac{d^k}{d\beta^k}\frac{1}{(\beta)_m},  \label{ii4}
\end{equation}
to represent the derivatives of the Pochhammer  and the reciprocal Pochhammer symbols with respect to their arguments.
Then, derivation with respect to $\varepsilon$ in Eq. (\ref{ii1}) gives, in view of Eq. (\ref{ii3bis}),
\begin{eqnarray}
\frac{\partial^n}{\partial \varepsilon^n}\ _p\! F_q \left( \!\! \left.\begin{array}{l}\alpha_1,\ldots,\alpha_p \\
\beta_1,\ldots,\beta_q\end{array}\right|z\right) &=& n!\,\sum_{m=0}^\infty\,\frac{z^m}{m!}\sum_{k_1=0}^n
a_1^{k_1}\,\mathcal{P}_m^{(k_1)}(\alpha_1)\nonumber \\
& & \hspace{-1cm}\times\,\sum_{k_2=0}^{n_1}a_2^{k_2}\,\mathcal{P}_m^{(k_2)}(\alpha_2)
\cdots\sum_{k_p=0}^{n_{p-1}}a_p^{k_p}\,\mathcal{P}_m^{(k_p)}(\alpha_p) \nonumber \\
& & \hspace{-2cm}\times \sum_{k_{p+1}=0}^{n_p}
b_1^{k_{p+1}}\,\mathcal{Q}_{m}^{(k_{p+1})}(\beta_1)\sum_{k_{p+2}=0}^{n_{p+1}}b_2^{k_{p+2}}
\,\mathcal{Q}_{m}^{(k_{p+2})}(\beta_2)\cdots  \nonumber \\
& & \hspace{-3cm}\cdots\sum_{k_{p+q-1}=0}^{n_{p+q-2}}b_{q-1}^{k_{p+q-1}}
\,\mathcal{Q}_{m}^{(k_{p+q-1})}(\beta_{q-1})\,b_q^{n_{p+q-1}}\,\mathcal{Q}_{m}^{(n_{p+q-1})}(\beta_{q}) \,.   \label{ii5}
\end{eqnarray}
The expression of $\mathcal{C}_n(z)$ then results immediately from the right-hand side of Eq. (\ref{ii5}), by suppressing the initial factor $n!$ and
substituting $A_i$ for $\alpha_i$ ($i=1, 2, \ldots, p$) and $B_j$ for $\beta_j$ ($j=1, 2, \ldots, q$). Simple explicit expressions for
$\mathcal{P}_m^{(k)}(A_i)$ and $\mathcal{Q}_{m}^{(k)}(B_j)$ are obtained in Appendices A and B, respectively. They are
\begin{eqnarray}
\mathcal{P}_0^{(k)}(A_i)&=& \delta_{k,0},  \qquad \mathcal{P}_m^{(k)}(A_i)=0 \quad \mbox{for}\quad k>m>0,   \label{ii7} \\
\mathcal{P}_m^{(k)}(A_i)&=& (-1)^{m-k}\sum_{l=0}^{m-k}(-1)^l\,{m \choose l}\,s(m\!-\!l, k)\,(A_i)_l \quad \mbox{for}\; m\geq k,
\label{ii8} \\
\mathcal{Q}_0^{(k)}(B_j)&=& \delta_{k,0}, \qquad  \mathcal{Q}_m^{(0)}(B_j)=1/(B_j)_m\,, \label{ii9} \\
\mathcal{Q}_{m}^{(k)}(B_j)&=& (-1)^{k}\,\sum_{l=0}^{m-1}\frac{(-1)^l}{l!\,(m-1-l)!}\,\frac{1}{(B_j+l)^{k+1}} \quad \mbox{for}\quad m,\,k>0,
\label{ii10}
\end{eqnarray}
The $s(n, k)$ in Eq. (\ref{ii8}) represent the Stirling numbers of the first kind \cite[\S 26.8]{nist}, whose generating relation is
\[
\left[\ln (1+x)\right]^k = k!\,\sum_{n=k}^\infty s(n, k)\,\frac{x^n}{n!}, \qquad   |x|<1.
\]

\subsection{Some lower parameters are singular}

Let us consider now the case of $\beta_1$, $\ldots$, $\beta_r$  ($0<r\leq q$) being singular because, for $j=1, \ldots, r$,
\begin{equation}
\quad B_j=-N_j, \qquad N_1\leq \ldots \leq N_r \quad \mbox{nonnegative integers.} \label{ii11}
\end{equation}
The series expansion in Eq. (\ref{ii1}) can be written in the form
\begin{eqnarray}
\ _p\! F_q \left( \!\! \left.\begin{array}{l}\alpha_1,\ldots,\alpha_{p}\\
\beta_1,\ldots,\beta_q\end{array}\right|z\right) &=& \sum_{m=0}^{N_1}\,\frac{(\alpha_1)_m \cdots(\alpha_p)_m}{(\beta_1)_m
\cdots(\beta_q)_m}\,\frac{z^m}{m!}  \nonumber \\
& & \hspace{-1cm}+\,\frac{\varepsilon^{-1}}{b_1}\sum_{m=N_1+1}^{N_2}\,\frac{(\alpha_1)_m \cdots(\alpha_p)_m}{\widehat{(\beta_1)}_m (\beta_2)_m
\cdots(\beta_q)_m}\,\frac{z^m}{m!}  \nonumber \\
& & \hspace{-2cm}+\,\frac{\varepsilon^{-2}}{b_1b_2}\sum_{m=N_2+1}^{N_3}\,\frac{(\alpha_1)_m
\cdots(\alpha_p)_m}{\widehat{(\beta_1)}_m\widehat{(\beta_2)}_m (\beta_3)_m \cdots(\beta_q)_m}\,\frac{z^m}{m!}+ \ldots \nonumber \\
& & \hspace{-3cm}\ldots +\frac{\varepsilon^{-r}}{b_1b_2\cdots b_r}\sum_{m=N_r+1}^{\infty}\,\frac{(\alpha_1)_m \cdots(\alpha_p)_m}{\widehat{(\beta_1)}_m
\cdots\widehat{(\beta_r)}_m(\beta_{r+1})_m\cdots (\beta_q)_m}\,\frac{z^m}{m!}\,,  \label{ii12}
\end{eqnarray}
where a sum is void if the lower limit of the summation index is larger than the upper one. We have represented by $1/\widehat{(\beta_j)}_m$ the
``regularized" reciprocal Pochhammer symbol
\begin{equation}
\frac{1}{\widehat{(\beta_j)}_m }\equiv \frac{b_j\,\varepsilon}{(\beta_j)_m}, \qquad \beta_j=-N_j+b_j\,\varepsilon, \quad 0\leq N_j<m,  \label{ii13}
\end{equation}
whose derivatives with respect to its variable,
\begin{equation}
\mathcal{\widehat{Q}}_{m}^{(k)}(\beta_j) \equiv \frac{1}{k!}\,\frac{d^k}{d\beta_j^k}\frac{1}{\widehat{(\beta_j)}_m}\,,    \label{ii14}
\end{equation}
are calculated in Appendix B. The result is
\begin{equation}
\mathcal{\widehat{Q}}_{m}^{(k)}(\beta_j) =(-1)^k \sum_{l=0,\, l\neq N_j}^{m-1}\frac{(-1)^l}{l!\,(m-1-l)!}\,\frac{N_j-l}{(\beta_j+l)^{k+1}}\,,
\label{ii15}
\end{equation}
and, for $\varepsilon=0$,
\begin{equation}
\left.\mathcal{\widehat{Q}}_{m}^{(k)}(\beta_j)\right|_{\varepsilon=0} = \mathcal{\widehat{Q}}_{m}^{(k)}(-N_j)=- \sum_{l=0,\, l\neq
N_j}^{m-1}\frac{(-1)^l}{l!\,(m-1-l)!}\,\frac{1}{(N_j-l)^{k}}\,,  \label{ii16}
\end{equation}
The $\varepsilon$-expansion is now of the form
\begin{equation}
\ _p\! F_q \left( \!\! \left.\begin{array}{l}\alpha_1,\ldots,\alpha_{p}\\
\beta_1,\ldots,\beta_q\end{array}\right|z\right) = \sum_{n=-r}^\infty\,\mathcal{C}_n(z)\,\varepsilon^n\,. \label{ii2}
\end{equation}
The coefficients $\mathcal{C}_n(z)$ are immediately obtained as a sum of the corresponding coefficients of the $\varepsilon$-expansions of the
sums in the right-hand side of Eq. (\ref{ii12}), which can be evaluated  by means of the algorithm used to obtain Eq. (\ref{ii5}) from Eq.
(\ref{ii1}).

\section{Some examples}

The procedure sketched above may be used to get algebraic expansions, when the parameters $A_i$ and $B_j$ take especially simple values. Nevertheless,
we find our method most suited to obtaining numerical expansions, for arbitrary values of the parameters, by means of an implementation of the procedure
in FORTRAN or \verb"Mathematica" or any similar language. The algorithm is so simple that almost any user could construct an efficient package. In this section we consider several examples for which results, obtained by different methods, are available in the literature. Our procedure is able to
reproduce the exact algebraic expansions, and it considerably improves the numerical ones.

\subsection{Algebraic expansion of a Gauss hypergeometric function}

As an example of the algebraic use of our algorithm, let us try to obtain the first terms of the $\varepsilon$-expansion of the Gauss hypergeometric
function
\begin{equation}
\ _2\! F_1 \left( \!\! \left.\begin{array}{l}a_1\,\varepsilon,\,a_2\,\varepsilon\\
1+b_1\,\varepsilon\end{array}\right|z\right).  \label{iii1}
\end{equation}
According to Eq. (\ref{ii5}) and its form for $\varepsilon =0$, the coefficients of the $\varepsilon$-expansion are given by
\begin{eqnarray}
\mathcal{C}_n(z)&=&\sum_{m=0}^\infty \frac{z^m}{m!}\,\sum_{k_1=0}^n a_1^{k_1}\,\mathcal{P}_m^{(k_1)}(A_1)\sum_{k_2=0}^{n-k_1}
a_2^{k_2}\,\mathcal{P}_m^{(k_2)}(A_2) \nonumber \\
& & \hspace{4cm}\times \,b_1^{n-k_1-k_2}\,\mathcal{Q}_{m}^{(n-k_1-k_2)}(B_1).  \label{iii2}
\end{eqnarray}
In the particular case of Eq. (\ref{iii1}), one has $A_1=0$, $A_2=0$, $B_1=1$. For these values of the parameters, Eqs. (\ref{ii7}) to (\ref{ii10})
reduce to
\begin{eqnarray}
\mathcal{P}_0^{(k)}(0)&=& \delta_{k,0},  \qquad \mathcal{P}_m^{(0)}(0)= \delta_{m,0}, \qquad  \mathcal{P}_m^{(k)}(0)=0 \quad\mbox{for}\quad k>m\,,
\label{iii3}  \\
\mathcal{P}_m^{(k)}(0)&=& (-1)^{m-k}\,s(m, k) \qquad \mbox{for}\quad m\geq k>0\,, \label{iii4} \\
\mathcal{Q}_0^{(k)}(1)&=& \delta_{k,0}, \qquad \mathcal{Q}_{m}^{(0)}(1)=1/m!, \quad \mathcal{Q}_{m}^{(1)}(1)=-\frac{1}{m!}\sum_{l=1}^m
\frac{1}{l}=-\frac{H_m}{m!}\,, \label{iii5} \\
\mathcal{Q}_{m}^{(k)}(1)&=& (-1)^{k}\sum_{l=1}^m \frac{(-1)^{l-1}}{l!\,(m-l)!}\,\frac{1}{l^k} \qquad \mbox{for}\quad m, k>0\,,  \label{iii6}
\end{eqnarray}
where one uses the harmonic number $H_m$:
\begin{equation}
H_m = \sum_{l=1}^m \frac{1}{l}\,.
\end{equation}

Then, the first coefficients of the $\varepsilon$-expansion are
\begin{eqnarray}
\mathcal{C}_0(z)&=& \sum_{m=0}^\infty \frac{z^m}{m!}\, \left(\mathcal{P}_m^{(0)}(0)\right)^2 \,\mathcal{Q}_{m}^{(0)}(1) \nonumber  \\
                &=& 1\,, \label{iii7}
\end{eqnarray}
\begin{eqnarray}
\mathcal{C}_1(z)&=&\sum_{m=0}^\infty\frac{z^m}{m!}\left[ b_1\,\left(\mathcal{P}_m^{(0)}(0)\right)^2\mathcal{Q}_{m}^{(1)}(1)+
(a_1+a_2)\,\mathcal{P}_m^{(0)}(0)\,\mathcal{P}_m^{(1)}(0)\,\mathcal{Q}_{m}^{(0)}(1)\right] \nonumber \\
&=&0\,,  \label{iii8}
\end{eqnarray}
\begin{eqnarray}
\mathcal{C}_2(z) &=& \sum_{m=0}^\infty\frac{z^m}{m!}\left[ b_1^2\,\left(\mathcal{P}_m^{(0)}(0)\right)^2\mathcal{Q}_{m}^{(2)}(1)
+ b_1(a_1+a_2)\,\mathcal{P}_m^{(0)}(0)\,\mathcal{P}_m^{(1)}(0)\,\mathcal{Q}_{m}^{(1)}(1)\right. \nonumber  \\
& & \hspace{20pt}+\,\left.\left( a_1a_2\left(\mathcal{P}_m^{(1)}(0)\right)^2+(a_1^2+a_2^2)\,\mathcal{P}_m^{(0)}(0)\,\mathcal{P}_m^{(2)}(0)\right)
\mathcal{Q}_{m}^{(0)}(1)\right] \nonumber \\
&=& \sum_{m=0}^\infty \frac{z^m}{m!} \,a_1\,a_2\,\left(s(m, 1)\right)^2\,\frac{1}{m!}, \label{iii9}
\end{eqnarray}
which, by using the representation of the Stirling number
\begin{equation}
s(m, 1)=(-1)^{m-1}\,(m-1)!,   \label{iii10}
\end{equation}
and the definition of the dilogarithm function, Li$_2(z)$ \cite[25.12.1]{nist}, can be written in the form
\begin{equation}
\mathcal{C}_2(z) = a_1\,a_2\, \mbox{Li}_2(z). \label{iii11}
\end{equation}
\begin{eqnarray}
\mathcal{C}_3(z) &=& \sum_{m=0}^\infty\frac{z^m}{m!}\,\left[b_1^3\left(\mathcal{P}_m^{(0)}(0)\right)^2\mathcal{Q}_{m}^{(3)}(1)
+ b_1^2(a_1\! +\! a_2)\,\mathcal{P}_m^{(0)}(0)\,\mathcal{P}_m^{(1)}(0)\,\mathcal{Q}_{m}^{(2)}(1)\right. \nonumber\\
& & \hspace{20pt}+\,b_1\left(a_1\,a_2\left(\mathcal{P}_m^{(1)}(0)\right)^2+(a_1^2+a_2^2)\,\mathcal{P}_m^{(0)}(0)\,\mathcal{P}_m^{(2)}(0)\right)
\mathcal{Q}_{m}^{(1)}(1)  \nonumber \\
&+&\left.\left( (a_1^2a_2\! +\! a_1a_2^2)\,\mathcal{P}_m^{(1)}(0)\,\mathcal{P}_m^{(2)}(0) +(a_1^3\, +\! a_2^3)
\,\mathcal{P}_m^{(0)}(0)\,\mathcal{P}_m^{(3)}(0)\right)\mathcal{Q}_{m}^{(0)}(1)\right]   \nonumber \\
&=& \sum_{m=1}^\infty\,\frac{z^m}{m!}\,b_1\,a_1\,a_2 \,\left(\mathcal{P}_m^{(1)}(0)\right)^2 \, \mathcal{Q}_{m}^{(1)}(1) \nonumber  \\
& &\hspace{80pt}  +\,\sum_{m=2}^\infty\,\frac{z^m}{m!}\,a_1\,a_2\,(a_1+a_2)\,\mathcal{P}_m^{(1)}(0)\,\mathcal{P}_m^{(2)}(0)\, \mathcal{Q}_{m}^{(0)}(1)
\nonumber \\
&=& a_1\,a_2\Bigg[b_1\sum_{m=1}^\infty\,\frac{z^m}{m!}\left(s(m, 1)\right)^2\, \left( -\frac{H_m}{m!}\right)\, \nonumber \\
& &\hspace{80pt}-\,(a_1+a_2)\,\sum_{m=2}^\infty\,\frac{z^m}{m!}\, s(m, 1)\,s(m, 2)\,\frac{1}{m!}\Bigg],   \label{iii12}
\end{eqnarray}
which, by using  the representations Eq. (\ref{iii10}) and
\begin{equation}
s(m, 2)=(-1)^{m}\,(m-1)!\,H_{m-1}  \label{iii14}
\end{equation}
of the Stirling numbers, can be written in the form
\begin{eqnarray}
\mathcal{C}_3(z)&=& a_1\,a_2\Bigg[-b_1\sum_{m=1}^\infty\,\frac{z^m}{m^2}\,H_m
+(a_1+a_2)\,\sum_{m=2}^\infty\,\frac{z^m}{m^2}\,H_{m-1}\Bigg]  \nonumber \\
&=& a_1\,a_2\Bigg[-b_1\sum_{m=1}^\infty\,\frac{z^m}{m^2}\frac{1}{m}
+(a_1+a_2-b_1)\,\sum_{m=2}^\infty\,\frac{z^m}{m^2}\,H_{m-1}\Bigg]. \label{iii15}
\end{eqnarray}
The last equation can be expressed in terms of the polylogarithm function, Li$_n(z)$ \cite[Eq. 25.12.10]{nist}, and Nielsen's generalized
polylogarithms, $S_{n,p}(z)$ \cite{kolb}, to give
\begin{equation}
\mathcal{C}_3(z) = a_1\,a_2\,\left[-b_1\,\mbox{Li}_3(z)+(a_1+a_2-b_1)\,S_{1,2}(z)\right)]. \label{iii16}
\end{equation}
We see that our method reproduces the results obtained by Kalmykov \cite[Eq. 4.7]{kalm} by the help of relations presented in Ref. \cite{flei}.

\subsection{Algebraic expansion of an Appell function}

Del Duca {\it et al.} \cite[Eq. (5.19)]{deld} have considered the algebraic $\varepsilon$-expansion of the Appell function
\begin{equation}
F_4 \left( \!\! \left.\begin{array}{l}1,\, 1+\varepsilon\\1+\varepsilon,\,1+\varepsilon\end{array}\right|x_1,\,x_2\right) \label{iii17}
\end{equation}
by using an algorithm based on the algebra properties of nested harmonic sums. From the series expansion of the Appell function
\begin{equation}
F_4 \left( \!\! \left.\begin{array}{l}\alpha_1,\, \alpha_2 \\ \beta_1,\,\beta_2\end{array}\right|x_1,\,x_2\right)=
\sum_{m_1=0}^\infty\sum_{m_2=0}^\infty
\frac{(\alpha_1)_{m_1+m_2}\,(\alpha_2)_{m_1+m_2}}{(\beta_1)_{m_1}\,(\beta_2)_{m_2}}\,\frac{x_1^{m_1}}{m_1!}\,\frac{x_2^{m_2}}{m_2!},
\label{iii18}
\end{equation}
application of our method would give for the coefficient of $\varepsilon^n$ in the $\varepsilon$-expansion the expression
\begin{eqnarray}
\mathcal{C}_n(x_1, x_2)&=&\sum_{m_1=0}^\infty \,\sum_{m_2=0}^\infty\,\frac{x_1^{m_1}}{m_1!}\,\frac{x_2^{m_2}}{m_2!}\,\sum_{k_1=0}^n
a_1^{k_1}\,\mathcal{P}_{m_1+m_2}^{(k_1)}(A_1)\sum_{k_2=0}^{n-k_1} a_2^{k_2}\,\mathcal{P}_{m_1+m_2}^{(k_2)}(A_2) \nonumber \\
& & \hspace{20pt}\times
\sum_{k_3=0}^{n-k_1-k_2}\,b_1^{k_3}\,\mathcal{Q}_{m_1}^{(k_3)}(B_1)\,b_2^{n-k_1-k_2-k_3}\,\mathcal{Q}_{m_2}^{(n-k_1-k_2-k_3)}(B_2),  \label{iii19}
\end{eqnarray}
which, for the values of the parameters in Eq. (\ref{iii17}) reduces to
\begin{eqnarray}
\mathcal{C}_n(x_1, x_2)&=&\sum_{m_1=0}^\infty \,\sum_{m_2=0}^\infty\,\frac{x_1^{m_1}}{m_1!}\,\frac{x_2^{m_2}}{m_2!}
\,\mathcal{P}_{m_1+m_2}^{(0)}(1)\sum_{k_2=0}^{n} \mathcal{P}_{m_1+m_2}^{(k_2)}(1) \nonumber \\
& & \hspace{80pt}\times \sum_{k_3=0}^{n-k_2}\,\mathcal{Q}_{m_1}^{(k_3)}(1)\,\mathcal{Q}_{m_2}^{(n-k_2-k_3)}(1),  \label{iii20}
\end{eqnarray}
By using Eqs. (\ref{ii7}) and (\ref{ii8}) with $A_i=1$ and Eqs. (\ref{iii5}) and (\ref{iii6}), we obtain for the first coefficients of the
$\varepsilon$-expansion (understanding that a sum is zero if the lower limit of the summation index is larger than the upper one)
\begin{eqnarray}
\mathcal{C}_0(x_1, x_2) &=& \sum_{m_1=0}^\infty \,\sum_{m_2=0}^\infty\,\frac{x_1^{m_1}}{m_1!}\,\frac{x_2^{m_2}}{m_2!}\,
\left(\mathcal{P}_{m_1+m_2}^{(0)}(1)\right)^2 \mathcal{Q}_{m_1}^{(0)}(1)\,\mathcal{Q}_{m_2}^{(0)}(1)  \nonumber \\
  &=& \sum_{m_1=0}^\infty  \sum_{m_2=0}^\infty\,{m_1+m_2 \choose m_1}^2\,x_1^{m_1}\,x_2^{m_2} \nonumber \\
  &=& F_4(1, 1; 1, 1; x_1, x_2).   \label{iii21}
\end{eqnarray}
\begin{eqnarray}
\mathcal{C}_1(x_1, x_2) &=& \sum_{m_1=0}^\infty \,\sum_{m_2=0}^\infty\,\frac{x_1^{m_1}}{m_1!}\,\frac{x_2^{m_2}}{m_2!}\, \mathcal{P}_{m_1+m_2}^{(0)}(1)
\Bigg[\mathcal{P}_{m_1+m_2}^{(1)}(1)\,\mathcal{Q}_{m_1}^{(0)}(1)\,\mathcal{Q}_{m_2}^{(0)}(1)   \nonumber \\
& & \hspace{20pt}+\,\mathcal{P}_{m_1+m_2}^{(0)}(1)\left(\mathcal{Q}_{m_1}^{(1)}(1)\,\mathcal{Q}_{m_2}^{(0)}(1)+
\mathcal{Q}_{m_1}^{(0)}(1)\,\mathcal{Q}_{m_2}^{(1)}(1)\right)\Bigg]  \nonumber \\
&=& \sum_{m_1=0}^\infty  \sum_{m_2=0}^\infty\,{m_1+m_2 \choose m_1}^2\,x_1^{m_1}\,x_2^{m_2} \left[ H_{m_1+m_2} - H_{m_1} - H_{m_2}\right]\,.
\label{iii22}
\end{eqnarray}
\begin{eqnarray}
\mathcal{C}_2(x_1, x_2) &=& \sum_{m_1=0}^\infty \,\sum_{m_2=0}^\infty\,\frac{x_1^{m_1}}{m_1!}\,\frac{x_2^{m_2}}{m_2!}\, \mathcal{P}_{m_1+m_2}^{(0)}(1)
\Bigg[\mathcal{P}_{m_1+m_2}^{(2)}(1)\,\mathcal{Q}_{m_1}^{(0)}(1)\,\mathcal{Q}_{m_2}^{(0)}(1)   \nonumber \\
& & \hspace{20pt}+\,\mathcal{P}_{m_1+m_2}^{(1)}(1)\Big(\mathcal{Q}_{m_1}^{(1)}(1)\,\mathcal{Q}_{m_2}^{(0)}(1)+
\mathcal{Q}_{m_1}^{(0)}(1)\,\mathcal{Q}_{m_2}^{(1)}(1)\Big)  \nonumber  \\
& & \hspace{-40pt}+\,\mathcal{P}_{m_1+m_2}^{(0)}(1)\Big(\mathcal{Q}_{m_1}^{(2)}(1)\,\mathcal{Q}_{m_2}^{(0)}(1)+
\mathcal{Q}_{m_1}^{(1)}(1)\,\mathcal{Q}_{m_2}^{(1)}(1)+\mathcal{Q}_{m_1}^{(0)}(1)\,\mathcal{Q}_{m_2}^{(2)}(1)\Big)\Bigg]  \nonumber \\
&=& \sum_{m_1=0}^\infty  \sum_{m_2=0}^\infty\,{m_1+m_2 \choose m_1}^2\,x_1^{m_1}\,x_2^{m_2}\Bigg[\sum_{l=2}^{m_1+m_2}\frac{H_{l-1}}{l} \nonumber  \\
& &\hspace{-60pt} -\,(-1)^{m_1+m_2}H_{m_1+m_2}
\left(H_{m_1}+H_{m_2}\right)+\left(\sum_{l=1}^{m_1}(-1)^{l-1}{m_1 \choose l}\frac{1}{l^2}\right) \nonumber  \\
& &\hspace{20pt}  +\,H_{m_1}H_{m_2}+\left(\sum_{l=1}^{m_2}(-1)^{l-1}{m_2 \choose l}\frac{1}{l^2}\right) \Bigg].  \label{iii23}
\end{eqnarray}
The same coefficients, with a different notation, have been obtained by Del Duca {\it et al.} \cite[Eq. (5.19)]{deld}. The reduction of the double sums (in $m_1$ and $m_2$) to known functions does not seem to be an easy task. In \cite[Eq. (5.90)]{deld}, transcendental functions $\mathcal{M}(\vec{i},
\vec{j}, \vec{k}; x_1, x_2)$ have been introduced to represent the different terms of the coefficients of the $\varepsilon$-expansion. The properties
of those $\mathcal{M}$ functions are discussed in an appendix of the same paper.

\subsection{Numerical expansion of several generalized hypergeometric functions}

As a first example of numerical application of our algorithm, we take
\begin{equation}
\ _4\! F_3 \left( \!\! \left.\begin{array}{l}-4\varepsilon, -\frac{1}{2}-\varepsilon, -\frac{3}{2}-2\varepsilon, \frac{1}{2}-3\varepsilon \\
-\frac{1}{2}+2\varepsilon, -\frac{1}{2}+4\varepsilon, \frac{1}{2}+6\varepsilon\end{array}\right|\frac{1}{2}\right), \label{iv1}
\end{equation}
whose $\varepsilon$-expansion has been considered by Carter and Heinrich \cite{cart} as a test for their \verb"SecDec" package and by Huang and Liu for their own \verb"NumExp" package \cite{huan}. Table 1 shows our results up to order $\varepsilon^{10}$, obtained by means of a pedestrian \verb"Mathematica" implementation of our method. Obviously, the $m$-sum in the series expansion of the hypergeometric function must be truncated at, say, $m=M$. The value of $M$ determines the accuracy of our results, which may be controlled by successive increments of $M$. Notice that, as all parameters are rational, \verb"Mathematica" gives, for the truncated $m$-sum, exact results in the form of rational numbers. The values shown in Table 1 are the decimal approximations of those rational results. For the parameters and variable in Eq. (\ref{iv1}) a value of $M=50$ is enough to give the digits reported. For comparison, we have calculated the first coefficients of the $\varepsilon$-expansion with the \verb"HypExp" package \cite{hub1,hub2}, which gives exact results in the case of integer or half-integer values of the parameters. We report, in the third column of Table  1, the numerical approximation, up to 15 significant digits, of those exact results. The computation to order $\varepsilon^6$ took 49 sec. An attempt of computation to order $\varepsilon^{10}$ required too much time and was considered unnecessary in view of the agreement of the first coefficients.
We present also, in the fourth column of Table 1, the expansion, up to $\varepsilon^{10}$, obtained by means of the numerical package \verb"NumExp" \cite{huan} with a step $e_h=10^{-6}$. The first coefficients are extremely precise: more than 20 correct significant digits (checked with our method truncated at $M=100$). But the precision decreases as the order in $\varepsilon$ increases. The authors of the package warn, in Ref. \cite{huan}, about this effect and point out its dependence on the size of $e_h$.
\begin{table}
\caption{Coefficients of the $\varepsilon$-expansion of the generalized hypergeometric function given in Eq. (\ref{iv1}).}
\begin{tabular}{lrcrcr}
\hline
$\varepsilon$ order & our method & \quad & HypExp & \quad & NumExp  \\
\hline
$\varepsilon^0$  & $1$ & & $1$ & & $1.00000000000000$  \\
$\varepsilon^1$  & $-4.27968776167886$ & & $-4.27968776167886$ & & $-4.27968776167886$ \\
$\varepsilon^2$  & $-26.6975474079466$ & & $-26.6975474079466$ & & $-26.6975474079466$ \\
$\varepsilon^3$  & $195.871193504205$ & & $195.871193504205$ & & $195.871193504205$ \\
$\varepsilon^4$  & $-7313.74176765086$ & & $-7313.74176765086$ & & $-7313.74176765086$ \\
$\varepsilon^5$  & $90693.2356441548$ & & $90693.2356441548$ & & $90693.2356441548$ \\
$\varepsilon^6$  & $-1426862.01660383$ & & $-1426862.01660383$ & & $-1426862.01660383$ \\
$\varepsilon^7$  & $17612046.1413323$ & & $17612046.1413323$ & & $17612406.1413322$ \\
$\varepsilon^8$  & $-233969019.148423$ & &  & & $-233969019.142915$ \\
$\varepsilon^9$  & $2846673719.75988$ & &  & & $2846673264.38359$ \\
$\varepsilon^{10}$ & $-35635855655.1898$ & &  & & $-35614562917.4105$ \\
\hline
\end{tabular}
\end{table}

In the second example,
\begin{equation}
\ _5\! F_4 \left( \!\! \left.\begin{array}{l}\varepsilon, -\varepsilon, -3\varepsilon, -5\varepsilon, -7\varepsilon \\ 2\varepsilon, 4\varepsilon,
6\varepsilon, 8\varepsilon\end{array}\right|\frac{1}{2}\right),   \label{iv2}
\end{equation}
we realize that all lower parameters are singular, as
\begin{equation}
B_1=B_2=B_3=B_4=0\,.
\end{equation}
Equation (\ref{ii12}) adopts then the form
\begin{equation}
\ _5\! F_4 \left( \!\! \left.\begin{array}{l}\alpha_1,\ldots,\alpha_5\\
\beta_1,\ldots,\beta_4\end{array}\right|z\right) = 1+\frac{\varepsilon^{-4}}{b_1b_2b_3b_4}\,\sum_{m=1}^{\infty}\,\frac{(\alpha_1)_m
\cdots(\alpha_5)_m}{\widehat{(\beta_1)}_m \cdots\widehat{(\beta_4)}_m}\,\frac{z^m}{m!},  \label{iv3}
\end{equation}
Nevertheless, this particular case is especially simple, as
\begin{equation}
A_1=A_2=A_3=A_4=A_5=0,
\end{equation}
and the negative exponent terms of the Laurent expansion become identically equal to zero. Equivalent to the expansion in Eq. (\ref{iv3}), we may write
the following one
\begin{equation}
\hspace{-25pt}\ _5\! F_4 \left( \!\! \left.\begin{array}{l}\alpha_1,\ldots,\alpha_5\\
\beta_1,\ldots,\beta_4\end{array}\right|z\right) = 1+\frac{a_1a_2a_3a_4a_5\,\varepsilon}{b_1b_2b_3b_4}\sum_{m=1}^{\infty}\frac{(1\!+\!\alpha_1)_{m-1}
\cdots(1\!+\!\alpha_5)_{m-1}}{(1\!+\!\beta_1)_{m-1} \cdots(1\!+\!\beta_4)_{m-1}}\,\frac{z^m}{m!},  \label{iv4}
\end{equation}
which can be treated as described in the case of no singular lower parameters. The algebraic procedure gives for the first coefficients of the
$\varepsilon$-expansion of
\[
\ _5\! F_4 \left( \!\! \left.\begin{array}{l}a_1\varepsilon,\ldots,a_5\varepsilon \\
b_1\varepsilon,\ldots,b_4\varepsilon\end{array}\right|z\right) \nonumber
\]
the expressions
\begin{eqnarray}
\mathcal{C}_0(z) &=& 1\,,  \nonumber  \\
\mathcal{C}_1(z) &=& \frac{a_1a_2a_3a_4a_5}{b_1b_2b_3b_4}\left(-\,\ln (1-z)\right)\,,  \nonumber  \\
\mathcal{C}_2(z) &=& \frac{a_1\cdots a_5}{b_1\cdots b_4}\left(a_1\!+\!\ldots\!+\!a_5\!-\!b_1\!-\!\ldots\! \!-\!b_4\right)\,\frac{1}{2}\,\left[\ln
(1-z)\right]^2\,, \nonumber
\end{eqnarray}
and so on. The numerical results of our method, for the values of the parameters and the variable given in Eq. (\ref{iv2}), are shown in Table 2,
together with those obtained by using \verb"HypExp" \cite{cart} and \verb"NumExp" \cite{huan}.
\begin{table}
\caption{Coefficients of the $\varepsilon$-expansion of the generalized hypergeometric function given in Eq. (\ref{iv2}).}
\begin{tabular}{lrcrcr} \hline
$\varepsilon$ order  & our method & \quad & HypExp & \quad & NumExp \\ \hline
$\varepsilon^0$  & $1$ & & $1$ & & $1.00000000000000$ \\
$\varepsilon^1$  & $0.189532432184360$ & & $0.189532432184360$ & & $0.189532432184360$ \\
$\varepsilon^2$  & $-2.29904274238202$ & &  $-2.29904274238202$ & & $-2.29904274238202$ \\
$\varepsilon^3$  & $55.4690190360554$ & & $55.4690190360554$ & & $55.4690190360554$ \\
$\varepsilon^4$  & $-1014.39242265234$ & & $-1014.39242265235$ & & $-1014.39242265235$ \\
$\varepsilon^5$  & $15729.382951742$ & & $15729.3829517422$ & & $15729.3829517422$ \\
$\varepsilon^6$  & $-216907.17756543$ & & $-216907.177565435$ & & $-216907.177565435$ \\
$\varepsilon^7$  & $2728106.3284185$ & & $2728106.32841847$ & & $2728106.32841847$ \\
$\varepsilon^8$  & $-31818216.953372$ & &   & & $-31818216.9529126$ \\
$\varepsilon^9$  & $348410894.51286$ & &   & & $348410894.565153$ \\
$\varepsilon^{10}$ & $-3617363078.8137$ & &   & & $-3615586412.42142$ \\
\hline \end{tabular} \end{table}

Huang and Liu \cite{huan} have proved the power of their method by applying it to the case of irrational parameters. Specifically, they have given the
expansion of
\begin{equation}
\ _4\! F_3 \left( \!\! \left.\begin{array}{l}-4\varepsilon, -\frac{1}{2}-\varepsilon, -\frac{\pi}{2}-2\varepsilon, \frac{1}{3}-3\varepsilon \\
-\pi+2\varepsilon, -\frac{1}{4}+4\varepsilon, \frac{1}{2}+6\varepsilon\end{array}\right|\frac{1}{2}\right). \label{iv5}
\end{equation}
Table 3 shows their results, together with ours.
\begin{table}
\caption{Coefficients of the $\varepsilon$-expansion of the generalized hypergeometric function given in Eq. (\ref{iv5}).}
\begin{tabular}{lrcr} \hline
$\varepsilon$ order  & our method & \quad & NumExp \\ \hline
$\varepsilon^0$ & $1$ & & $1.00000000000000000$ \\
$\varepsilon^1$ & $-1.44555526747928$ & & $-1.44555526747927565$ \\
$\varepsilon^2$ & $3.938387944727$ & & $3.93838794472745744$ \\
$\varepsilon^3$ & $-266.9473544234$ & & $-266.947354423423669$ \\
$\varepsilon^4$ & $298.666582673$ & & $298.666582668478365$ \\
$\varepsilon^5$ & $-56037.4042903$ & & $-56037.4029214013816$ \\
$\varepsilon^6$ & $-113001.082396$ & & $-113205.384759634797$ \\
\hline \end{tabular} \end{table}

\section{Final comments}

We have presented a direct and intuitive procedure to construct the $\varepsilon$-expansion of any function, of one or several variables, of the
hypergeometric class, whenever the function can be written as a convergent power series of the variables. The building blocks are the successive
derivatives of Pochhammer and reciprocal Pochhammer symbols particularized for the values of the parameters at $\varepsilon=0$. We have provided
closed explicit expressions for those derivatives to any order, although recurrence relations, given in Appendices A and B, can equally be used to
evaluate them.  Each coefficient of the $\varepsilon$-expansion appears as a power series of the variable or variables that can be summed, to give a
known function, only in very special cases. This fact, however, is not a serious drawback, in our opinion. In fact, the symbols commonly used to refer
to most of the special functions are merely a shorthand to represent a series or an integral. Nevertheless, what makes our procedure especially
useful is its easy implementation to obtain numerical values of the coefficients. The examples given above show that our method allows one to calculate terms of higher order with a notable accuracy. Besides this, the proposed method is self-contained: it does not need auxiliary procedures such as, for
instance, differential reduction, even in the case of singular lower parameters.

An important limitation to the use of the algorithm presented in this paper stems from the necessary convergence of the series representation of the
function to be expanded in powers of $\varepsilon$. This restricts the applicability of the method to certain domains of the variables. Obviously, one
may have recourse to analytic continuation. For instance, in the very common case of $\ _{q+1}\! F_q$ one could use, for $|z|>1$, the relation
\cite[Eq. (10)]{huan} \cite[Sec. 5.3, Eq. (3)]{luke}
\begin{eqnarray}
\ _{q+1}\! F_q \left( \!\! \left.\begin{array}{l}\alpha_1,\ldots,\alpha_{q+1} \\ \beta_1,\ldots,\beta_q\end{array}\right|z\right) &=&
\frac{\Gamma(\beta_1)\cdots\Gamma(\beta_q)}{\Gamma(\alpha_1)\cdots\Gamma(\alpha_{q+1})} \sum_{h=1}^{q+1}\frac{\Gamma(\alpha_h)\prod_{l=1;l\neq
h}^{q+1}\Gamma(\alpha_l-\alpha_h)}{\prod_{j=1}^q\Gamma(\beta_j-\alpha_h)} \nonumber \\
& & \hspace{-60pt}\times\,(z^{-1}e^{i\pi})^{\alpha_h}\ _{q+1}\! F_q \left( \!\! \left.\begin{array}{l}\alpha_h,\{1+\alpha_h-\beta_k\}_{k=1, \ldots, q}
\\ \{1+\alpha_h-\alpha_k\}_{k=1, \ldots, q+1; k\neq h} \end{array}\right|\frac{1}{z}\right).
\label{v1}
\end{eqnarray}
Our procedure also gets in trouble when the variable (or variables), being inside the convergence domain, lies (lie) near the border. Although the resulting series of powers of the variable are convergent, they may be useless, due to their slow convergence, for a numerical computation of the coefficients of the $\varepsilon$-expansion. The difficulty may be overcome in the case of the Gauss hypergeometric function, for which linear and quadratic transformations of the variable are available \cite[Sec. 15.8]{nist} \cite[Eqs. 15.3.3 to 15.3.32]{abra}. In the case of $\ _{q+1}\! F_q $ with $q>1$, instead, the solutions about the singular point $z=1$ are not of hypergeometric type, and analogous transformations of the variable lead to complicated expressions. The issue has been addressed by N{\o}rlund \cite{norl}. Transformations of the variables for Appell functions can be found in Sec. 16.16 of Ref. \cite{nist}.
The discussion of the analytic continuation of all functions of the hypergeometric class is a hard task, and is out of the scope of this paper.

\section*{Acknowledgments}

We thank E. de Rafael  and G. Vulvert for their comments on the manuscript. The suggestions of two anonymous referees have contributed to a considerable improvement in the presentation of this paper. The work has been supported by the Spanish DGIID-DGA grant 2009-E24/2, the Spanish  MICINN grants FPA2009-09638 and CPAN-CSD2007-00042 and by Departamento de Ciencia, Tecnolog\'{\i}a y Universidad del Gobierno de Arag\'on (Project
E24/1) and Ministerio de Ciencia e Innovaci\'on (Project MTM2009-11154).

\appendix

\section{Derivatives of the Pochhammer symbol $(\alpha)_m$ with respect to its argument $\alpha$}

We introduced, in Eq. (\ref{ii4}), the notation
\begin{equation}
\mathcal{P}_m^{(k)}(\alpha) \equiv \frac{1}{k!}\,\frac{d^k}{d\alpha^k}(\alpha)_m\,,    \label{a1}
\end{equation}
As $(\alpha)_m$ is a polynomial of degree $m$ in $\alpha$,
\begin{equation}
\mathcal{P}_m^{(k)}(\alpha) = 0 \qquad\mbox{for}\qquad k>m .  \label{a2}
\end{equation}
We assume in the rest of this appendix that $m\geq k$.
A generating function for the $\mathcal{P}_m^{(k)}(\alpha)$ can be immediately obtained from \cite[Section 6.2.1, Eq. (2)]{luke}
\begin{equation}
\sum_{m=0}^\infty (\alpha)_m (-t)^m/m! \equiv \ _1\!F_0(\alpha;;-t) = (1+t)^{-\alpha}, \qquad |t|<1. \label{a3}
\end{equation}
Derivation, $k$ times, with respect to $\alpha$ gives
\begin{equation}
\sum_{m=0}^\infty k!\, \mathcal{P}_m^{(k)}(\alpha)\, (-t)^m/m! = (-1)^k\,(1+t)^{-\alpha}\, \left[\ln(1+t)\right]^k, \qquad |t|<1. \label{a4}
\end{equation}
Then, we have
\begin{eqnarray}
\hspace{-1cm}\mathcal{P}_m^{(k)}(\alpha) &=& \frac{(-1)^{k-m}}{k!}\,\left.\frac{\partial^m}{\partial t^m}\left((1+t)^{-\alpha}\,
\left[\ln(1+t)\right]^k\right)\right|_{t=0}  \nonumber \\
&=& \frac{(-1)^{k-m}}{k!}\,\sum_{l=0}^m {m \choose l}\,\left.\left(\frac{\partial^l}{\partial t^l}(1+t)^{-\alpha}\right)\left(\frac{d^{m-l}}{d
t^{m-l}}\left[\ln(1+t)\right]^k\right)\right|_{t=0}\hspace{-12pt}. \label{a5}
\end{eqnarray}
Obviously,
\begin{equation}
\left.\frac{\partial^l}{\partial t^l}(1+t)^{-\alpha}\right|_{t=0} = (-1)^l\,(\alpha)_l.  \label{a6}
\end{equation}
On the other hand, as $[\ln (1+t)]^k$ is the generating function of the Stirling numbers \cite[Eq. 26.8.8]{nist},
\begin{equation}
[\ln (1+t)]^k = k! \sum_{n=k}^\infty s(n, k)\, t^n/n!, \qquad |t|<1,  \label{a7}
\end{equation}
we get
\begin{equation}
\left.\left(\frac{d^{m-l}}{d t^{m-l}}[\ln(1+t)]^k\right)\right|_{t=0} = k!\,s(m-l, k)\,.  \label{a8}
\end{equation}
Substitution of Eqs. (\ref{a6}) and (\ref{a8}) in (\ref{a5}) gives
\begin{equation}
\mathcal{P}_m^{(k)}(\alpha) = (-1)^{m-k}\sum_{l=0}^{m-k}(-1)^l{m \choose l}\,s(m-l, k)\,(\alpha)_l \qquad\mbox{for}\quad m\geq k. \label{a9}
\end{equation}

For numerical computation, however, it may be preferable to make use of the recurrence relation
\begin{equation}
\mathcal{P}_{m+1}^{(k)}(\alpha)=(\alpha+m)\,\mathcal{P}_m^{(k)}(\alpha)+\mathcal{P}_m^{(k-1)}(\alpha), \qquad k>0, \label{a10}
\end{equation}
with starting values
\begin{equation}
\mathcal{P}_{0}^{(k)}(\alpha)=\delta_{k,0}, \qquad \mathcal{P}_m^{(0)}(\alpha)=(\alpha)_m. \label{a11}
\end{equation}

\section{Derivatives of the reciprocal Pochhammer symbol $1/(\beta)_m$ with respect to its argument $\beta$}

The notation
\begin{equation}
\mathcal{Q}_{m}^{(k)}(\beta) \equiv \frac{1}{k!}\,\frac{d^k}{d\beta^k}\frac{1}{(\beta)_m}  \label{b1}
\end{equation}
was proposed in Eq. (\ref{ii4}) to represent the derivatives of the reciprocal Pochhammer symbol with respect to its variable. Very simple
expressions for these $\mathcal{Q}_{m}^{(k)}(\beta)$ can be easily obtained from the relation \cite[Eq. 4.2.2.45]{prud}
\begin{equation}
 \frac{1}{(\beta)_m} = \sum_{l=0}^{m-1}\frac{(-1)^l}{l!\,(m-1-l)!}\,\frac{1}{\beta+l}\,, \qquad m>0\,.   \label{b2}
\end{equation}
Direct derivation with respect to $\beta$ in this equation gives
\begin{equation}
\mathcal{Q}_{m}^{(k)}(\beta)= (-1)^k \sum_{l=0}^{m-1}\frac{(-1)^l}{l!\,(m-1-l)!}\,\frac{1}{(\beta+l)^{k+1}}\,, \qquad m>0\,.  \label{b3}
\end{equation}
provided that $\beta$ is different from a nonpositive integer, $-N$, such that $0\leq N<m$. To deal with this case (in the limit $\varepsilon\to 0$), we defined, in Eq. (\ref{ii13}),  the regularized reciprocal Pochhammer symbol
\begin{equation}
\frac{1}{\widehat{(\beta)}_m }\equiv \frac{b\,\varepsilon}{(\beta)_m}, \qquad \beta=-N+b\,\varepsilon, \quad 0\leq N<m.  \label{b4}
\end{equation}
whose derivatives were denoted, in Eq. (\ref{ii14}), by
\begin{equation}
\mathcal{\widehat{Q}}_{m}^{(k)}(\beta) \equiv \frac{1}{k!}\,\frac{d^k}{d\beta^k}\frac{1}{\widehat{(\beta)}_m}    \label{b5}
\end{equation}
From Eqs. (\ref{b2}) and (\ref{b4}) we obtain
\begin{equation}
\frac{1}{\widehat{(\beta)}_m }=\delta_{m, 1}+\sum_{l=0}^{m-1}\frac{(-1)^l}{l!\,(m-1-l)!}\,\frac{N-l}{\beta+l}.  \label{b6}
\end{equation}
Repeated derivation with respect to $\beta$ leads to
\begin{equation}
\mathcal{\widehat{Q}}_{m}^{(k)}(\beta) =(-1)^k \sum_{l=0,\, l\neq N}^{m-1}\frac{(-1)^l}{l!\,(m-1-l)!}\,\frac{N-l}{(\beta+l)^{k+1}}\,, \qquad k>0\,.
\label{b7}
\end{equation}

For numerical implementation, one may use the recurrence relations
\begin{equation}
\mathcal{Q}_{m+1}^{(k)}(\beta)=\left(\mathcal{Q}_{m}^{(k)}(\beta) - \mathcal{Q}_{m+1}^{(k-1)}(\beta)\right)/(\beta+m), \label{b8}
\end{equation}
with initial values
\begin{equation}
\mathcal{Q}_0^{(k)}(\beta)=\delta_{k,0}, \qquad   \mathcal{Q}_{m}^{(0)}(\beta)=1/(\beta)_m, \label{b9}
\end{equation}
and, for the regularized reciprocal Pochhammer symbols when $\beta=-N+b\,\varepsilon$,
\begin{equation}
\widehat{\mathcal{Q}}_{m+1}^{(k)}(\beta)=\left(\widehat{\mathcal{Q}}_{m}^{(k)}(\beta) - \widehat{\mathcal{Q}}_{m+1}^{(k-1)}(\beta)\right)/(\beta+m),
\qquad m>N\,,\label{b10}
\end{equation}
starting with
\begin{equation}
\widehat{\mathcal{Q}}_{N+1}^{(k)}(\beta)=\mathcal{Q}_N^{(k)}(\beta), \qquad   \widehat{\mathcal{Q}}_{m}^{(0)}(\beta)=1/\widehat{(\beta)}_m,  \qquad
m>N\,. \label{b11}
\end{equation}

\end{document}